\documentclass[aps,prb,twocolumn,showpacs,superscriptaddress]{revtex4-2}
\usepackage{graphicx}
\usepackage{siunitx}
\usepackage{xcolor}
\usepackage{amsmath,amssymb}
\usepackage[T1]{fontenc}
\usepackage{lmodern}
\usepackage[utf8]{inputenc}
\usepackage[english]{babel}
\usepackage{natbib}
\usepackage{hyperref}
\hypersetup{colorlinks=true,citecolor={blue},linkcolor={blue},urlcolor={blue}}

\newcommand{\hsn}[1]{{\color{black}{#1}}}

\newcommand{\ag}[1]{{\color{black}{#1}}}

\newcommand{\cn}[1]{{\color{black} #1}}

\newcommand{\new}[1]{{\color{black} #1}}

\newcommand{\neww}[1]{{\color{black} #1}}

\begin{document}
\title{Reconfigurable quantum fluid molecules of bound states in the continuum}

\author{Antonio Gianfrate}
\affiliation{CNR Nanotec, Institute of Nanotechnology, via Monteroni, 73100, Lecce, Italy}

\author{Helgi Sigur{\dh}sson}
\email{helg@hi.is}
\affiliation{Science Institute, University of Iceland, Dunhagi-3, IS-107 Reykjavik, Iceland}
\affiliation{Institute of Experimental Physics, Faculty of Physics, University of Warsaw, ul.~Pasteura 5, PL-02-093 Warsaw, Poland}

\author{Vincenzo Ardizzone}
\affiliation{CNR Nanotec, Institute of Nanotechnology, via Monteroni, 73100, Lecce, Italy}

\author{Hai Chau Nguyen}
\affiliation{Naturwissenschaftlich-Technische Fakult\"at, 
Universit\"at Siegen, Walter-Flex-Stra{\ss}e 3, 57068 Siegen, Germany}

\author{Fabrizio Riminucci}
\affiliation{Molecular Foundry, Lawrence Berkeley National Laboratory,
One Cyclotron Road, Berkeley, California, 94720, USA}

\author{Maria Efthymiou-Tsironi}
\affiliation{CNR Nanotec, Institute of Nanotechnology, via Monteroni, 73100, Lecce, Italy}

\author{Kirk W. Baldwin}
\affiliation{PRISM, Princeton Institute for the Science and Technology of Materials, Princeton University, Princeton, New Jersey 08540, USA}

\author{Loren N. Pfeiffer}
\affiliation{PRISM, Princeton Institute for the Science and Technology of Materials, Princeton University, Princeton, New Jersey 08540, USA}

\author{Dimitrios Trypogeorgos}
\affiliation{CNR Nanotec, Institute of Nanotechnology, via Monteroni, 73100, Lecce, Italy}

\author{Milena De Giorgi}
\affiliation{CNR Nanotec, Institute of Nanotechnology, via Monteroni, 73100, Lecce, Italy}

\author{Dario Ballarini}
\email{dario.ballarini@nanotec.cnr.it}
\affiliation{CNR Nanotec, Institute of Nanotechnology, via Monteroni, 73100, Lecce, Italy}

\author{Hai Son Nguyen}
\affiliation{Univ Lyon, Ecole Centrale de Lyon, INSA Lyon, Universit\'e  Claude Bernard Lyon 1, CPE Lyon, CNRS, INL, UMR5270, Ecully 69130, France}
\affiliation{Institut Universitaire de France (IUF), 75231 Paris, France}

\author{Daniele Sanvitto}
\affiliation{CNR Nanotec, Institute of Nanotechnology, via Monteroni, 73100, Lecce, Italy}

\date{\today}

\begin{abstract}

\neww{Topological bound states in the continuum are confined wave-mechanical objects that offer advantageous ways to enhance light-matter interactions in compact photonic devices. In particular, their large quality factor in the strong-coupling regime has recently enabled the demonstration of Bose-Einstein condensation of bound-state-in-the-continuum polaritons. Here, we show that condensation into a negative-mass bound state in the continuum exhibits interaction-induced state confinement, opening opportunities for optically reprogrammable molecular arrays of quantum fluids of light. We exploit this optical trapping mechanism to demonstrate that such molecular complexes show hybridization with macroscopic modes with unusual topological charge multiplicity. Additionally, we demonstrate the scalability of our technique by constructing extended mono- and diatomic chains of bound-state-in-the-continuum polariton fluids that display non-Hermitian band formation and the opening of a minigap. Our findings offer insights into large-scale, reprogrammable, driven, dissipative many-body systems in the strong-coupling regime.}

\end{abstract}

\maketitle

\section{Introduction}
 Development of artificially structured materials at the nanoscale in solid-state technologies and photonics has opened opportunities to explore the coupling between matter and confined light. In the recent years, confined photonic modes---known as symmetry protected bound states in the continuum (BICs)---have emerged in full force through precise parameter tuning in synthetic lattices across various wave-mechanical and quantum systems~\cite{Hsu_NatRevMat2016}. In optical systems~\cite{Azzam_AdvOptMat2021}, these extremely high $Q$-factor BIC resonances cannot radiate energy despite residing within the light cone, opening perspectives on ultralow threshold lasing devices emitting coherent vectorial light~\cite{Kodigala_Nature2017, Hwang_NatComm2021} associated with the inherent momentum-space topological charge of BICs~\cite{Zhen_PRL2014, Doeleman_NatPho2018}. Aside from low-threshold lasing devices, photonic BICs have a wide reaching application from plasmonics to photonic waveguides and crystals including sensing~\cite{Chen_OptLett2022}, filtering~\cite{Foley_PRB2014}, and enhancing light-matter interactions~\cite{Aigner_Arxiv2022}. Other interesting optical effects like multistability~\cite{Krasikov_PRB2018} and solitons~\cite{Dolinina_PRE2021} have been predicted for BICs when nonlinearity is introduced to the system. 
 
Quite recently, BICs operating in the strong light-matter coupling regime in subwavelength grated semiconductor heterostructures were proposed~\cite{Lu_PhotRes2020} and demonstrated~\cite{Kravtsov_LSA2020, dang_realization_2022}. In these structures emergent light-matter bosonic quasiparticles known as exciton-polaritons (here after polaritons) can undergo nonequilibrium Bose-Einstein condensation~\cite{Byrnes_NatPhys2014} in a low-loss BIC mode~\cite{Ardizzone_Nature2022}.

Polaritons are formed due to the strong coupling of light (photons) and matter (excitons) in semiconductor cavities with embedded quantum wells~\cite{Deng_RMP2010}. The photonic component gives polaritons extremely small effective mass, whereas the exciton component makes them highly interactive~\cite{Ciuti_PRB1998}. Upon condensation, the macroscopic polariton wavefunction is explicitly encoded in the spontaneous coherent emission of photons from the cavity, giving direct access to all of its degrees of freedom through standard optical techniques. Various phenomena associated with other branches of physics have now been demonstrated in polariton condensates including topological insulators~\cite{Klembt_Nature2018}, quantized vorticity and superfluidity~\cite{Sanvitto_NatPhys2010}, spontaneous phase-~\cite{Berloff_NatMat2017, Tao_NatMat2022} as well as pseudospin-pattern formation~\cite{Ohadi_PRL2017}, and much more.

\begin{figure*}[t]
    \centering
    \includegraphics[width=0.95\linewidth]{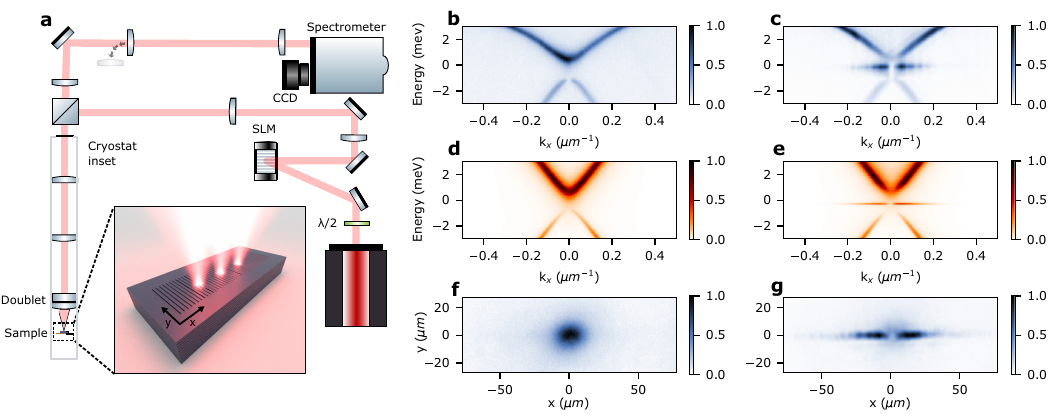}
    \caption{{\bf The exciton-polariton BIC condensate in a quantum-well grating waveguide}: (a) Schematic of the experimental setup. In the excitation line the surface of the spatial light modulator (SLM) is recreated in the Fourier plane of the excitation objective. According to the requirements of the particular experiment, the sample's luminescence is captured by a 2f detection line, and either the real or Fourier plane is reconstructed on the monochromator entrance.     
    The inset is an artistic representation of the grated waveguide sample. (b),(c) Experimental energy momentum dispersion crosscut along $k_{y} =0$ below and above condensation threshold, respectively. (d),(e) Corresponding calculated dispersions using a Dirac equation (see Eq.~\eqref{eq.dirac} and Methods). Energy scales are shifted with respect to the energy of the gap state in (c). (f),(g) Corresponding measured real space distribution of the emission.}
    \label{fig1}
\end{figure*}

Considerable effort \cn{in contemporary photonics} has been dedicated \cn{to} designing tailored polariton potential landscapes~\cite{Schneider_RPP2016} for simulation of exotic Hamiltonians~\cite{Amo_CRPhys2016} and possibly unconventional computing purposes~\cite{Kavokin_NatRevPhys2022}. The engineered potential typically stems from irreversible techniques, requiring new samples for different tasks, based on e.g., etching~\cite{Galbiati_PRL2012}, patterning mesas~\cite{Kaitouni_PRB2006} or polymer layers~\cite{Jayaprakash_ACSPho2020}, or metallic deposition~\cite{Kim_NatPhys2011} on the microcavity surface. Alternatively, structured nonresonant excitation beams acting on the exciton component of the polariton have the advantage of being fully reversible giving also the possibility of dynamical fast modification of the potential landscape. The nonresonant beam photoexcites a co-localized reservoir of hot excitons which locally amplify and blueshift polaritons~\cite{Wertz_NatPhys2010}, underpinning the phenomena of ballistically coupled polariton Bose-Einstein condensates~\cite{Alyatkin_NatComm2021}, optical trapping~\cite{Cristofolini_PRL2013, Askitopulos_PRB2013}, and all-optical polariton lattices~\cite{Pickup_NatComm2020, Alyatkin_NatComm2021, pieczarka_topological_2021}. When the beam is removed, the reservoir rapidly decays and, subsequently, so does the pump-induced polariton potential, underlining the reprogrammable feature of the system.

For conventional cavity polaritons with positive effective mass the pump-induced localized blueshift means that outside of the pump region condensate polaritons can convert elastically into states with high $k$-vectors with subsequent fast expansion velocities. This means that polaritons are always repelled from their gain region, increasing the condensate threshold power and lowering its coherence time. The contrary, an excitation profile which provides effective local redshift would form an attractive potential in the gain region. This implies that polaritons can be optically confined while still being efficiently pumped, a scheme that has not been explored properly up to date.

Here, we \ag{empower} this new paradigm by using negative mass BIC polaritons in subwavelength \hsn{quantum-well waveguide gratings}~\cite{Ardizzone_Nature2022}.
\neww{Until now, the negative mass scheme has primarily been explored in polaritonics using micropillar arrays \cite{Baboux_Optica2018,tanese_polariton_2013} with several fundamental differences to our BIC platform.
First, our waveguide approach overcomes a geometric limitation of the micropillars, enabling sub-micrometer patterning without the exciton quenching issues.
Second, the diffracting coupling mechanism and losses between polaritonic branches and the continuum can be engineered giving control over the BIC properties and non-trivial topology.
}

\neww{In this work we }\ag{demonstrate how BIC polariton condensates allow us to optically and reversibly structure macroscopically coherent nonlinear fluids of light} \ag{to simulate} molecular bonding between different optical traps. We \ag{provide experimental realization of} coherent quantum mode hybridization in a BIC condensate dyad (two pump spots) similar to a bosonic Josephson junction, Bloch band formation in an artificial mono-atomic quasi-1D chain (ten spots), and subband formation with celebrated minigap opening in staggered chains analogous to the Su-Schrieffer-Heeger (SSH) model. Our results are the first evidence of using nonlinearly spatially-localized BICs to design extended evanescently coupled many-body systems that preserve the vectorial momentum-space topological textures. Interestingly, our results can be approximately described with a massive \ag{non-Hermitian} Dirac model due to the linear dispersion of waveguided photons \ag{opening a new pathway in exploring synthetic many-body Dirac Hamiltonians in the strong light-matter coupling regime}. \ag{Furthermore} adding nonlinearities to BICs we show that it leads to surprisingly effective spatial confinement of these very long-lived states. 
\cn{This is in sharp contrast with the conventional BICs \hsn{in photonic lattices}, which are delocalized states at least along one dimension~\cite{Hsu_NatRevMat2016}.}

\section{Results}
\subsection{Single trapped BIC condensate}
Our sample consists of a planar semiconductor waveguide with a grating along the $x$-direction and embedded GaAs-based quantum wells (see Fig.~\ref{fig1}a and Methods). The presence of the grating folds the guided photonic modes lying outside the light cone across $k_x=0$~\cite{Lu_PhotRes2020}. An anticrossing occurs due to finite non-Hermitian coupling between counterpropagating modes through the radiative continuum with a gap opening at the $\Gamma$-point, directly tunable through the grating filling factor and etching depth~\cite{Riminucci2022}. By tuning the phase of this complex-valued coupling it is possible to entirely suppress the losses for one of the photonic eigenstates at the $\Gamma$-point, corresponding to a symmetry protected BIC. In our case, the BIC is located in the lower energy branch (see Fig.~\ref{fig1}b). \hsn {Namely, the two counter-propagating polariton guided modes are coupled via the gratings (i.e. diffractive coupling) and also via the radiative continuum (i.e. radiative coupling). The effects of these coupling is two-fold: i) gap opening  with two polaritonic branches, and ii) destructive interference of radiative losses for the lower polariton branch at the normal incidence, leading to a nodal line in the far-field emission.}

\begin{figure*}[t]
    \centering
    \includegraphics[width=\linewidth]{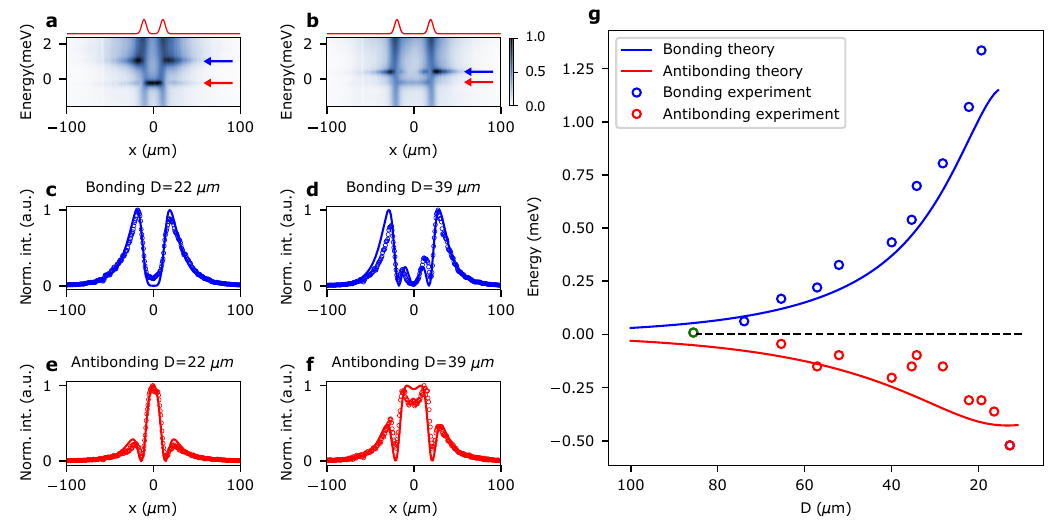}
    \caption{{\bf BIC molecule condensate  characterisation:} (a,b) Energy resolved real space PL crosscut at $y=0$ showing the double trap case for an pump spot separation distance of 39 and \SI{22}{\micro\metre}, respectively. \new{The red lines are a scaled schematic representation showing the pump profile.} (c-f) Experimental (markers) and theoretical (solid lines) mode profiles of the hybridized trapped condensates. (g) Energies of the hybridized modes as a function of pump spot separation distance for constant power. Dashed line refers to the isolated condensate energy $E_0$ (shifted to zero for brevity $E = \Bar{E}-E_{0}$ with $E_{0} = 1520.2~meV $). Blue and red markers represent the upper (\hsn{bonding}) and lower (\hsn{antibonding}) mode energies and corresponding solid lines represent the energies from the Dirac model. The green marker indicate\hsn{s} the measured energy for \hsn{single trapped condensate}.}
    \label{fig2}
\end{figure*}

As the quantum well excitons become strongly coupled with the guided photons\cn{,} the resultant polaritons inherit the BIC state from the photons and strong inter-particle Coulomb interactions from the excitons. The unique combination of these properties has recently enabled ultralow threshold polariton condensation under nonresonant excitation~\cite{Ardizzone_Nature2022}. We demonstrate this effect in Fig.~\ref{fig1} for a single Gaussian pump spot (see Methods for details on the experiment). 
\hsn{Figures~\ref{fig1}b and~\ref{fig1}d show, respectively, the measured and modeled far field photoluminescence (PL) having in-plane wavevector along the corrugation direction.} 
This scheme excites polaritons nearly uniformly across the two dispersion branches below the exciton energy~\cite{note1}. The BIC at the $\Gamma$-point of the lower branch does not radiate into the continuum, as expected, which manifests as a dark notch. 

When the pump is increased above condensation threshold the emission changes dramatically (see Figs.~\ref{fig1}{c,e})~\cite{Ardizzone_Nature2022}. The long-lifetime BIC results in enhanced amalgamation polaritons at the lower branch extremum through spontaneous scattering. At high enough powers, stimulated scattering is triggered into the BIC state where a negative effective mass polariton condensate forms. The pump induced blueshift then attracts polaritons to the spot, effectively confining the condensate. This is in contrast to ballistic polariton condensates in planar cavities where most of the condensate converts into large momentum outflowing polaritons~\cite{Alyatkin_NatComm2021}. The fundamental mode of the pump-induced trap is shifted into the bandgap and is visibly occupied by polaritons as seen in Figs.~\ref{fig1}{c,e}. The corresponding \hsn{PL measured in real space} is shown in Fig.~\ref{fig1}{g}. Notice that the fundamental trap mode emits light with an odd-parity (i.e., a dark nodal line at $k_x=0$) \new{and that the incoherent PL in the upper branch suffers a slight depletion around $k_x\approx0$ because of the localized pumping}.
\begin{figure*}[t]
    \centering
    \includegraphics[width=\linewidth]{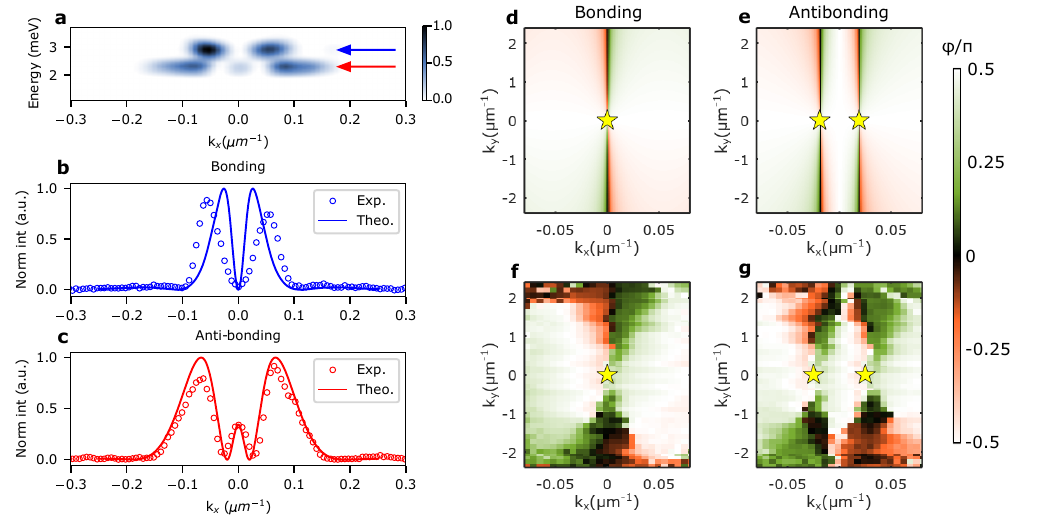}
    \caption{{\bf Momentum space photoluminescence characteristics in the \hsn{double-spot arrangement}:   } (a) Energy-momentum crosscut along $k_y= 0$, for a double trap potential with the pumps separated by \SI{31}{\micro\metre}. The two arrows mark the energies of the \hsn{bonding} and \hsn{antibonding} \ag{above thresholds, i.e. for higher excitation power.} \ag{Here the condensates signals results blueshifted respect to Fig~\ref{fig2} and no residual photoluminescence from the non condensed states can be observed.} (b,c) Corresponding experimental (markers) and calculated (solid curves) PL line profiles of the two states. Calculated (d,e) and measured (f,g) momentum space distribution of the condensate polarization vector $\phi$ for the \hsn{bonding} and \hsn{antibonding} state, respectively.}
    \label{fig3a}
\end{figure*}

\neww{At low momentum, and for sufficiently negative exciton-photon detuning, roughly exceeding half the light-matter Rabi coupling strength, the polariton dispersion can be approximated by a simplistic single-particle model. A non-Hermitian 1D Dirac Hamiltonian describing the coupling between massless forward and backward propagating polariton modes $\psi_{F,B}$ with velocity $v$  along the waveguide~\cite{Lu_PhotRes2020, Ardizzone_Nature2022},}
\begin{equation} \label{eq.dirac}
   \hat{H} = \begin{pmatrix}
   -i\hbar v \partial_x -i\gamma  + V(x) & U +i\gamma \\
   U+i\gamma & i \hbar v \partial_x - i\gamma  + V(x)
   \end{pmatrix}.
\end{equation}
Here, the pump induced exciton reservoir is taken as a static potential $V(x)$ because of the exciton's much heavier mass.
%
%
%
The propagating polaritons are coupled through $U$ coming from the periodic corrugation of the waveguide and---being folded above the light line---each of them leak to the radiate continuum with the same coupling strength $\gamma$~\cite{Lu_PhotRes2020} (note, we have neglected non-radiative exciton losses which are not essential in our results). \hsn{The radiative channel makes it possible for a loss exchange mechanism of the same strength $\gamma$. Such an interference via radiating waves dictates the far-field emission of polariton modes and is the origin of the polariton BIC formation.}  
\new{In absence of the potential $V(x)$, the eigenmodes of the Hamiltonian (\ref{eq.dirac}) in $k$-space give rise to two polaritonic bands with opposite curvature $\pm v^2/2U$. These bands are separated by a gap of $2U$. The loss exchange, on the other hand, affects the mode losses, resulting in a mode with a vanishing linewidth, i.e. the BIC state, and a mode that absorbs all the radiative losses \cite{Ardizzone_Nature2022}.}
The blueshifting potential is taken to be Gaussian $V(x) = V_0 e^{-x^2/2w^2}$ of width $w$. When the pump is weak then $V_0 \approx 0$ and we recover the non-trapped polariton dispersion shown in Fig.~\ref{fig1}{b,d} (see Methods for details on calculations). When the pump is strong ($V_0>0$) we see a clear mode within the bandgap in agreement with experiment corresponding to the fundamental mode of the effective confining potential $V(x)$ (see Fig.~\ref{fig1}{c,e}).

\begin{figure*}[t!]
    \centering
    \includegraphics[width=\linewidth]{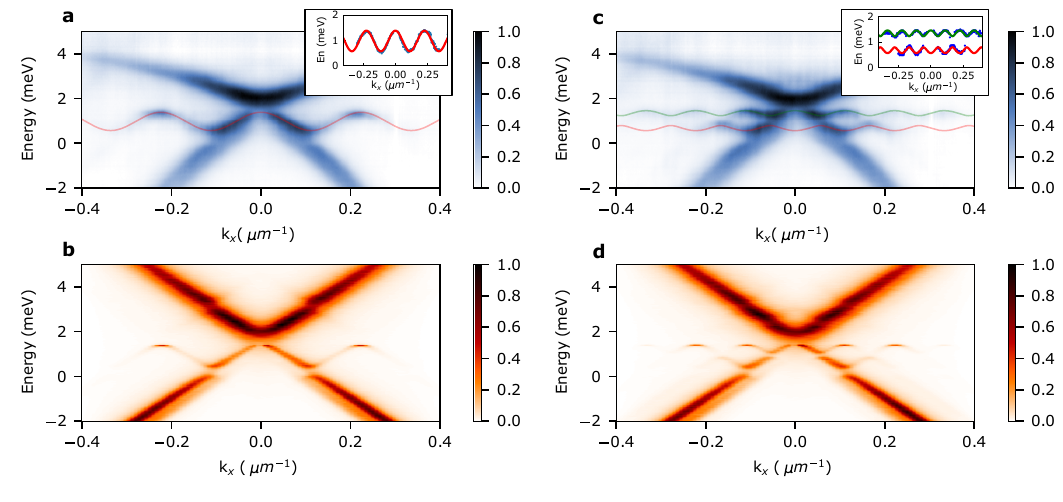}
    \caption{{\bf 10-BIC condensate chain} (a) Experimental energy momentum crosscut showing the additional band for an uniform 10 spot chain. Here the PL contribution of the exciton has been subtracted to better highlight the dispersion. Red curve shows the tight binding approximation for the $s$-band, $E = 2J \cos{(k_x a)}$. (b) Corresponding calculated dispersion using the lossy Dirac model~\eqref{eq.dirac}. (c) Measured energy-momentum PL crosscut for a staggered lattice with a staggering contrast $a/b = 1.75$. The red and green lines depict the fitting of the SSH dispersion $E_{1,2} = \pm \sqrt{J_a^2 + J_b^2 + 2 J_a J_b \cos{[(a+b)k_x]}}$. The inset show the resulting curves overlapped with the extracted experimental data. The fit parameters are $J_{a}=0.33$, $J_{b}=0.11$ meV, $a=$\SI{20}{\micro\metre}, $b=$\SI{36}{\micro\metre}. (d) Calculated dispersion of a staggered lattice using the parameters extracted from the measure in panel c. In this panel the energy is re-scaled to the energy of the not blueshifted BIC state as $E= \Bar{E}-E_{0}$ with $E_{0}=1524.8$~meV.
    }
    \label{fig3}
\end{figure*}

\subsection{BIC polariton molecules}
To demonstrate the novel concept of our trapped condensates, we structure our excitation beam into two spatially separated spots thus creating a double trap potential or alternatively a driven-dissipative bosonic Josephson junction. This forms a very simple setting to study fundamental quantum mechanical effects such as mode hybridization but now with added BIC topological charge structure in $k$-space. This is in contrast to conventional double-trap studies (i.e., no BIC) using ultracold atomic gases~\cite{Gati_IOP2007}, other polariton systems~\cite{Lagoudakis_PRL2010, Galbiati_PRL2012, Abbarchi_NatPhys2013} and photonic condensates~\cite{Kurtscheid_Science2019}.

The condensate energy-resolved near-field PL in the double trap is shown in Figs.~\ref{fig2}{a,b} for trap separation of $D=39$ and \SI{22}{\micro\metre}\cn{,} respectively. Mode hybridization can be clearly observed for both cases\cn{,} as the condensate starts populating the new \hsn{bonding} and \hsn{antibonding} BIC states characterized by clear normal-mode splitting. The corresponding spatial line-profile envelopes of the \hsn{bonding} and \hsn{antibonding} modes are shown in Fig.~\ref{fig2}{c,d} and~\ref{fig2}{e,f} with overlaid analytical modes obtained \hsn{ from solving the bound states of the equation ~\eqref{eq.dirac} with $V(x)$ having the form of a double Gaussian potential, given by $V(x)=V_0 e^{-(x-D/2)^2/2w^2} + V_0 e^{-(x+D/2)^2/2w^2}$ (see Methods) }. Because of the characteristic of BIC, light emitted from the waveguide depicts an unconventional arrangement of nodes (corresponding to $\pi$ phase jumps) in the total condensate wavefunction that are captured by our model.
Differently from previous polariton schemes, we can continuously tune the evanescent inter-condensate coupling and, consequently, the energy splitting between the bonding and antibonding states by simply adjusting the spot separation distance $D$ as shown in Fig~\ref{fig2}{g}. \hsn{As expected, decreasing the distance between the two spots enhances the overlap between the trapped condensates, exponentially splitting the bonding and anti-bonding states}. 
\cn{Interestingly, the blue-shift of the bonding state increases much faster than the red-shift of the anti-bonding one when decreasing the distance $D$. In particular, when $D<$\SI{16}{\micro\metre}, trap contains only the anti-bonding state (see Fig.~\neww{S1} in the Supplemental Information for details). Indeed, for small distance $D$, the blue shift of the bonding-state is so strong that it falls into the energy band of the structure outside the trap. These available states allow polaritons to tunnel out of the traps via the Klein tunnelling mechanism\,\cite{Allain2011}.}
\hsn{The bound state energies computed from the theoretical model~\eqref{eq.dirac}, despite its simplicity, reproduces the measurement data to a fairly good accuracy with a single fit parameter $V_0=3.5$ meV; all other parameters are fixed from experiments carried out below the condensation threshold.}
\hsn{The far-field profiles in Figs~\ref{fig2}{c-f} were then computed without any further fitting (see Methods).} 

In order to fully characterize the BIC molecules, we studied their topological structure by investigating the condensate momentum space polarization. BICs are characterised by topological charges~\cite{Zhen_PRL2014} corresponding to polarization vortices in momentum space \neww{(Fig. S2)}. It was recently confirmed that a single polariton BIC condensate could indeed inherit this topological structure~\cite{Ardizzone_Nature2022}. However, there has been no investigation into the topological structure of extended BIC condensates, or BIC molecules. Figure~\ref{fig3a}{a} shows the energy resolved cross section of the momentum space PL showing the molecule \hsn{bonding} and \hsn{antibonding} states \ag{above threshold.} \ag{The} corresponding comparison between the experimental and theoretical reciprocal space line profiles are depicted in Fig.~\ref{fig3a}{b,c} showing good agreement.\new{We attribute the small discrepancy in panel \ref{fig3a}{b} to the dynamical/transient blueshift of the state because the experiment is performed under femtosecond pulsed excitation (see Methods and Supplemental Video 1 ). }
Here again the phase jumps can be clearly distinguished as dark nodes. Performing an energy and polarization resolved tomography we retrieved the momentum space maps of the polarization direction $\varphi$ (i.e., orientation of the polarization ellipse). \hsn{The polarization textures were also theoretically calculated from the eigenvectors of Dirac equation \eqref{eq.dirac} (see Methods)}. The resulting theoretical and experimental maps for the bound and anti-bound state are shown in Fig.~\ref{fig3a}{d,e} and Fig.~\ref{fig3a}{f,g}, respectively.
These maps show the presence of a single polarization vortex in the bound state and a pair of vortices in the anti-bound state. This finding underlines that each node in the hybridized double-trap system corresponds to a polarization vortex having the same topological charge of the original photonic BIC. Therefore, the topological charge of the hybridized state has increased with respect to the modes of the uncoupled traps, a finding unreported to date.

By further increasing the complexity of the potential landscape (by increasing the number of traps), it is possible to add more and more states within the gap with alternating parity. This is clearly illustrated in Fig.~S3 in the Supplemental Material which compares single, double, and triple trap configurations.

\subsection{BIC condensate chains}
Lastly, we studied the feasibility of simulating large-scale systems by coupling 10 condensates together arranged in a finite 1D chain with a lattice constant $a$. Figures~\ref{fig3}{a} and~\ref{fig3}{b} show experimental and calculated energy-resolved PL in momentum space for a regular (i.e. effectively mono-atomic) 10 spot chain extracted along the $x$ direction.
Polaritons, within their lifetime, experience the discrete translational symmetry in the bulk of the chain, giving rise to Bloch modes and associated crystalline bands, the hallmark of solid-state physics. Approximating the polaritons as deeply confined in their pump-induced traps we can fit the observed Bloch band with the standard expression for mono-atomic ($s$-orbital) dispersion from tight-binding theory, $E = 2J \cos{(k_x a)}$ where $J$ is the coupling (hopping) energy between sites (red curve in Fig.~\ref{fig3}{b} with $a=$\SI{28}{\micro\metre} and $J = 0.2$ meV). The goodness of the fit underlines that the optically trapped polaritons here can be accurately described using tight-binding theory in contrast to coupled ballistic condensates~\cite{Pickup_NatComm2020} \neww{(Please refer to Fig. S3 for a direct comparison between the two mechanisms)}.

Another notable information that can be extracted from this experiment is related to the Bloch band topology. At $k_x = 0$ the Bloch modes inherit the same topology of the BIC mode as demonstrated by the lack of emission at the $\Gamma$-point. However for the off $\Gamma$ Brillouin zones the notch is absent\neww{, as also confirmed by the numerical simulations presented in Fig. S4}. 

To further demonstrate the advantage of our reconfigurable all-optical lattice we show in Fig.\ref{fig3}(c,d) the dispersion for a staggered lattice with a contrast $a/b=1.75$. Here, we observe the opening of a minigap as two distinct Bloch bands form, as expected \neww{from tight binding considerations}.
The green and red curves in in Fig. \ref{fig3}{c} represent the classical dispersion for a dimer lattice $E_{1,2} = \pm \sqrt{J_a^2 + J_b^2 + 2 J_a J_b \cos{[(a+b)k_x]}}$, with $J_a$ and $J_b$ the hopping coefficients and $a$ and $b$ the site distances. This type of a staggered system is a polariton analogue of the SSH Hamiltonian which is perhaps one of the simplest models to possess topological nontrivial gap opening and associated protected edge states. The full characterisation of these artificial polariton lattice systems is beyond the scope of the current study. We however predict that a wealth of nonlinear phenomena such as solitons and persistent Bloch oscillations can be studied in our evanescently coupled optical lattices offering a contrasting viewpoint with respect to ballistic lattices~\cite{Pickup_NatComm2020} and adding an extra control-knob on the non-Hermitian character of each "monomer".

\section{Discussion}
We have shown that it is possible to optically construct macroscopically coherent artificial molecules and atomic chains using polariton condensates in an extreme nonequilibrium setting protected from the continuum. For this purpose we have used a subwavelength \hsn{grating} waveguide with embedded quantum wells \hsn{possessing} a photonic \hsn{BIC} which, through strong light-matter coupling, enhances the lifetime of emergent polaritons and allows them to condense at the extremum of a negative mass dispersion \new{around $k_x \approx 0$, opening a path to explore driven nonequilibrium negative-mass hydrodynamics~\cite{Khamehchi_PRL2017}. This is in stark contrast to the $|\mathbf{k}_\parallel|>0$ extrema in the so-called anomalous lower polariton dispersion which cannot display condensation~\cite{Wurdack_NatComm2023}}. Because of their strong interactions, the negative mass polaritons become self-trapped leading to efficient condensation co-localised with the pump spot. 

As a result, a structured light source can reconfigurably write-in various potential landscapes \cn{that are} accurately described through evanescently coupled trapped in-plane waves and tight binding models reminiscent of optical lattice for cold atoms~\cite{Schafer_NatRevPhys2020}. 
This technique overcomes a severe challenge in realizing reprogrammable macroscopic many-body systems with continuously tunable parameters in the strong light-matter coupling regime~\cite{Schneider_RPP2016} with exciting prospects for many-body polariton simulation~\cite{Kavokin_NatRevPhys2022} \new{such as driven macroscopic quantum self-trapping and Josephson oscillations~\cite{Abbarchi_NatPhys2013} and quantum-dissipative phase transitions in optically reconfigurable non-Hermitian tight-binding lattices~\cite{Hartmann_NatPhys2006}}. Indeed, the self-trapping of our waveguided polaritons is in strong contrast to the ballistic non-trapped condensate polaritons in conventional and fabrication costly planar Bragg reflector cavities~\cite{Pickup_NatComm2020, Alyatkin_NatComm2021}. Here, no sample post-processing to create potential landscapes \cite{st-jean_lasing_2017} is required, and our reprogrammable optical-trapping technique efficiently stimulates polaritons in the trap (i.e., the optical gain is inside the trap) in contrast to other methods where the gain is outside the trap~\cite{pieczarka_topological_2021}. This constitutes a significant novelty in the solid state condensation panorama opening new possibilities in investigating complex many-body Hamiltonians in a continuously re-configurable system like demonstrated in Fig.~\ref{fig2}.

A remarkable observation is the increase of momentum space topological charges when two or more BIC condensates are brought together to hybridize into extended modes as shown in Fig.~\ref{fig3a}. \neww{As more condensates are added, the number of accessible topological charges in principle increases but at the cost of becoming more spread out in the far field due to the decreasing contrast between antinodal lines in the near field.} This finding could lead to controllable generation of polarization vortices complementing the surging interest in utilizing optical vortices and vectorial coherent light sources for communication and information processing technologies~\cite{Jincheng_Science2021}.

When multiple pump spots are brought together (see Fig.~\ref{fig3}) our system becomes described by Bloch's theorem and we gain access into non-Hermitian lattice Hamiltonians. This is particularly exciting from the perspective of being able to simulate many fundamental crystalline systems by exploiting the strong polariton interactions while at the same time being able to optically read out all the relevant dynamics. For this purpose we have demonstrated tunable transition from a mono-atomic 1D chain to the Su-Schrieffer-Heeger chain~\cite{Atala_NatPhys2013} in  Fig.~\ref{fig3} with clear minigap opening and subband formation. 

We have also shown that our results can be reasonably reproduced using a 1D Dirac Hamiltonian with a mass term. \hsn{The Dirac Hamiltonian appears in many low dimensional systems such as graphene, transition-metal dichalcogenides, and around the crossing of spin-bands in a Rashba Hamiltonian. It describes intriguing electron transport effects around corresponding Dirac cones giving rise to quantum Hall physics, topological phases, and Weyl semimetals. Its appearance in BIC polariton condensates holds promises for further exploration into nonlinear driven-dissipative Dirac dynamics. } Recently, Dirac cones have gained a great deal of interest in photonic~\cite{Lu_NatPho2013, Li_LightSciAppl2021, Krol_NatComm2022} and polariton systems~\cite{Milicevic_PRX2019, Polimeno_Optica2021} to bring associative topological electron concepts into the field of topological photonics. An interesting perspective of this Dirac Hamiltonian, for a future study, is that the \hsn{grating} filling factor can be adjusted to change the sign of the mass parameter $U$~\cite{Lu_PhotRes2020}. This implies that an interface between two gratings with different sign of $U$ is described by the Jackiw-Rebbi model~\cite{Jackiw_PRD1976, Tran_PRA2017, Lee_Nanopho2021} possessing a zero-energy midgap state. 

\section{Acknowledgements}
H.S. acknowledges the project No. 2022/45/P/ST3/00467 co-funded by the Polish National Science Centre and the European Union Framework Programme for Research and Innovation Horizon 2020 under the Marie Skłodowska-Curie grant agreement No. 945339; and the Icelandic Research Fund (Rannis), Grant No. 239552-051.
A.G, V.A., D.T., M.D., D.B., D.S acknowledge 
the Italian Ministry of University (MUR) for funding through the PRIN project “Interacting Photons in Polariton Circuits” – INPhoPOL (grant 2017P9FJBS),
the project “Hardware implementation of a polariton neural network for neuromorphic computing”–Joint Bilateral Agreement CNR-RFBR (Russian Foundation for Basic Research)–Triennal Program 2021–2023, 
the MAECI project “Novel photonic platform for neuromorphic computing”, Joint Bilateral Project Italia-Polonia 2022-2023, 
PNRR MUR project: "National Quantum Science and Technology Institute"- NQSTI ( PE0000023), 
PNRR MUR project: "Integrated Infrastructure Initiative in Photonic and Quantum Sciences" - I-PHOQS (IR0000016),
and the project FISR - C.N.R. “Tecnopolo di nanotecnologia e fotonica per la medicina di precisione” - CUP B83B17000010001 and ”Progetto Tecnopolo per la Medicina di precisione, Deliberazione della Giunta Regionale n. 2117 del 21/11/2018. 
H.S.N is funded by the French National Research Agency (ANR) under the project POPEYE (ANR-17-CE24-0020) and the IDEXLYON from Université de Lyon, Scientific Breakthrough project TORE within the Programme Investissements d’Avenir (ANR-19-IDEX-0005). He is also supported by the Auvergne-Rh\^{o}ne-Alpes region in the framework of PAI2020 and the Vingroup Innovation
Foundation (VINIF) annual research grant program under Project Code VINIF.2021.DA00169.
H.C.N. acknowledges the Deutsche Forschungsgemeinschaft (DFG, German Research Foundation, project numbers 447948357 and 440958198), the Sino-German Center for Research Promotion (Project M-0294), and the ERC (Consolidator Grant 683107/TempoQ). This research is funded in part by the Gordon and Betty Moore Foundation’s EPiQS Initiative, grant GBMF9615 to L.P., and by the National Science Foundation MRSEC grant DMR 2011750 to Princeton University. Work at the Molecular Foundry is supported by the Office of Science, Office of Basic Energy Sciences, of the U.S. Department of Energy under Contract No. DE-AC02-05CH11231. We thank Scott Dhuey for assistance with electron beam lithography and Paolo Cazzato for the technical support.

\section{Author contribution statement}

AG performed the experiments and the data analysis with the support of MET and VA. HS, VA and AG edited the manuscript with the input of all the authors. LNP, KWB, and FR fabricated and post processed the sample. HS, HSN and HCN provided the theoretical framework and reproduced the experimental results through numerical simulation. DT and MDG provided insight on the physical processes and helped in the data interpretation. DT provided the code to control the setup. DB and DS supervised the work.

\section{Competing interests statement}
The authors declare no competing interests.

\section{data availability statement}

The raw experimental data and the code used in this study are available from the corresponding author upon reasonable request.

\section{Methods}
\subsection{Sample and experiment}
A 500 nm thick Al$_{0.4}$Ga$_{0.6}$As planar waveguide embedded with $12$ GaAs quantum wells, 20 nm thick and spaced apart by 20 nm serves as the sample for our experiments \cite{Ardizzone_Nature2022, SurezForero2020, Riminucci2022}, \neww{ the resulting exciton transitions are investigated in Fig. S6}. The waveguide heterostructure is etched over an $300 \times$\SI{50}{\micro\metre} area along the $x$ and $y$ direction, respectively, such to possess a one-dimensional (1D) grating along the $x$-direction. Several replicas of the grating are etched on the sample with slightly different periods and filling factors, in order to finely tune the dispersion properties~\cite{Riminucci2022}, \neww{refer to Fig. S7 for the effect of the grating pitch on to the polariton dispersion}.

\new{It is important to note that an ideal BIC state can only exist in infinite structures. However, the finite size of the grating and the strong coupling with the excitonic transition impose an upper limit on the BIC lifetime, effectively resulting in the polariton BIC being a quasi-BIC state. A quantitative estimation of the quality factor can be found in Section S8 and S9 of the Supplemental Material.}

In order to prevent exciton dissociation and maintain strong light-matter coupling, the sample is cooled to 4 K during the experiment using a closed loop helium cryostat. The excitation is performed non-resonantly using an 80 Mhz, fs pulsed laser at 770 nm wavelength. The laser profile focused on the sample is precisely shaped with a spatial light modulator. A feedback method on the collected PL from the sample is employed to ensure uniformity across multiple spots~\cite{Topfer_Optica2021}. The spot size for all the results displayed in this manuscript is set to \SI{6.5}{\micro\metre} full-width-at-half-maximum (FWHM). All results in the study are extracted along the center of the waveguide (i.e., along $k_x$ at $k_y=0$).

We note that despite the effective attractive pull of the pump spot onto the polaritons the real space PL in Fig.~\ref{fig1}{e} extends much further, $\approx$\,\SI{20}{\micro\metre} , than the \SI{6.5}{\micro\metre} FWHM of the excitation spot \neww{(see Fig. S10)}. This is due to the enhanced diffusion of excitons sustaining the condensate which obtain a larger group velocity in the strong coupling regime.

\subsection{Modeling}
Polaritons in our systems are dictated by the Dirac Hamiltonian \eqref{eq.dirac} in the basis of counter-propagating polaritons of group velocity $\pm v$. The Hamiltonian includes the diffractive coupling of strength $U$, the radiative loss/coupling of strength $\gamma$ and the potential $V(x)$ from the excitonic reservoir that is induced by the structured optical pump. For multispot excitation configuration of $N$ spots focused at $x_{j=1..N}$, each spot induces a Gaussian potential of height $V_0$ and waist $w$. The potential $V(x)$ is then given by $V(x)=V_0\sum_{j=1}^{N}{e^{-(x-x_j)^2/2w^2}}$. 

For all \hsn{theoretical} calculations we \hsn{use} $w =$ \SI{8.5}{\micro\metre} corresponding to \SI{20}{\micro\metre} FWHM trap size. \hsn{This trap size value is extracted from the real-space photoluminescence measurement (see Fig.~S5 in the Supplemental Information). As explained previously} the larger trap size with respect to the pump \SI{6.5}{\micro\metre} FWHM is due to the finite diffusion of low-momentum excitons from the pump spot which gain substantial group velocity in the strong coupling regime.  The loss parameter is taken as $\gamma = 0.153$ meV based on past results~\cite{Ardizzone_Nature2022}. \hsn{The values of $U$ and $v$ are directly extracted from the polariton dispersion below threshold exhibiting a band gap of $2U$ and curvature $\pm \frac{v^2}{2U}$ (see Fig.~S1 in the Supplemental Information for measured dispersions below threshold). The experimental measurement have been performed on three  grating structures having three different detuning: Grating 1 ($U=0.7$ meV, $v=$\SI{32}{\micro\metre} ps$^{-1}$) for single trapped BIC condensate experiment, Grating 2 ($U=1.2$ meV, $v=$\SI{42}{\micro\metre} ps$^{-1}$) for BIC polariton molecules experiment and Grating 3 ($U=0.4$ meV, $v=$\SI{24}{\micro\metre} ps$^{-1}$) for BIC condensate chains.}\cn{ Therefore, the only fitting parameter for all theoretical calculation is $V_0=3.5$ meV.} 

The calculated dispersions in Figs.~\ref{fig1}{d,e} and Fig.~\ref{fig3}{b} were obtained by averaging over the dynamics of multiple random initial conditions in \eqref{eq.dirac} (i.e., Monte-Carlo sampling) \new{that were weighted by the pump profile $V(x)$. The depletion seen around $k_x\approx0$ in the simulated PL in Fig.~\ref{fig1}(e) can be attributed to the resonant coupling of blueshifted low momentum fluctuations in the upper branch to $k>0$ momentum states}. \hsn{The energy of molecule states in Fig.~\ref{fig2}{g} and their profiles in Fig.~\ref{fig2}{c-f} and Fig.~\ref{fig3}{b,c} are obtained by numerically solving the bound state solutions of the Dirac equation \eqref{eq.dirac}. Specifically, the obtained near-field profiles (in both real and momentum space) are converted into far-field profiles by suppressing the Fourier components corresponding to guided modes below the light cone. Finally, the polarization patterns in Fig.~\ref{fig3}{d,e} are calculated from the spinor components of the Dirac equation solutions. Detailed theoretical framework for Dirac polaritons will be discussed elsewhere.}

\bibliography{biblio}

\end{document}


\title{Supplementary information: Reconfigurable quantum fluid molecules of bound states in the continuum}

\date{\today}
\maketitle

\begin{figure*}[t]
    \centering
    \includegraphics[width=0.8\linewidth]{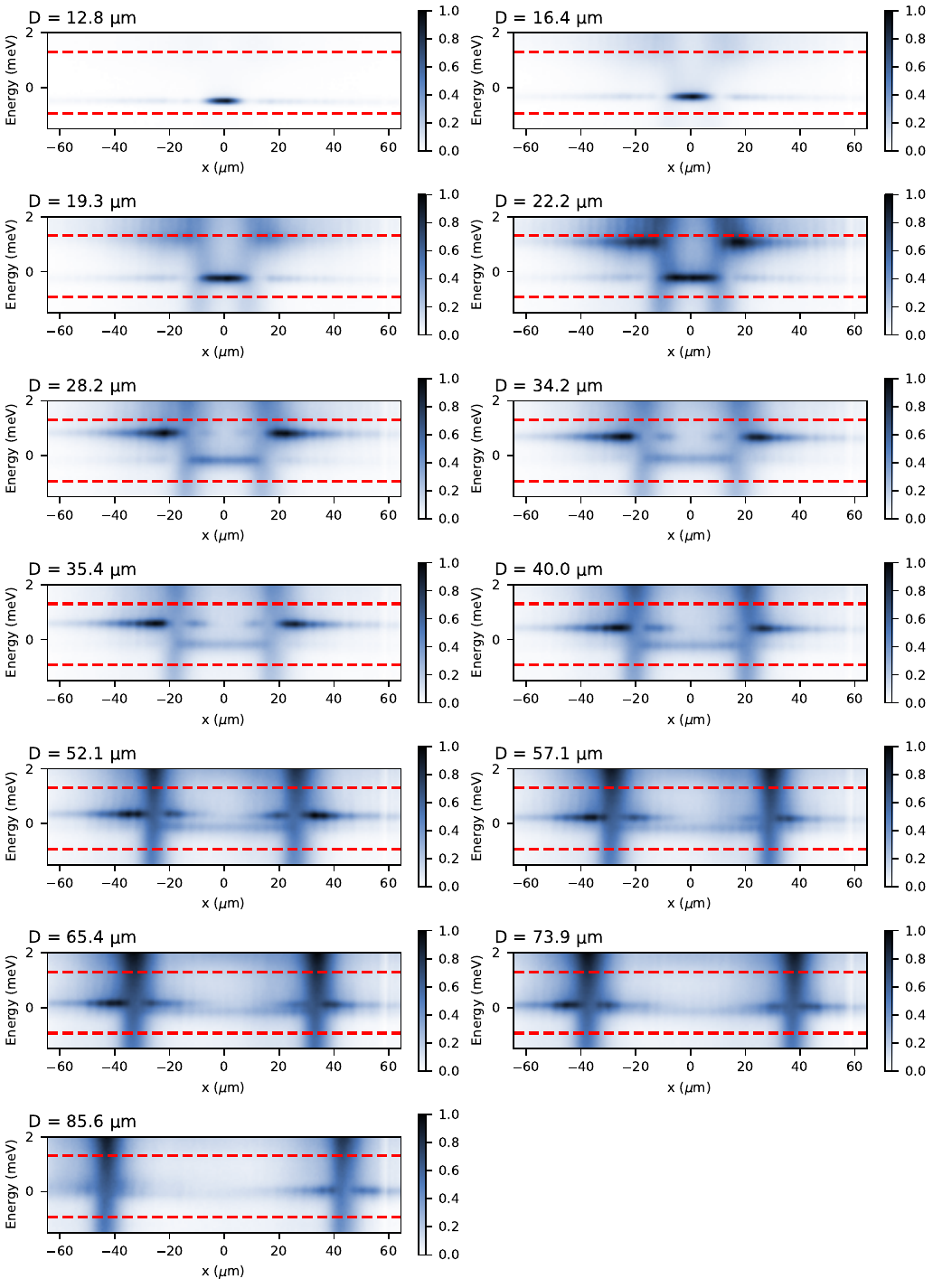}
    \caption{{\bf Experimental bonding and antibonding state splitting as a function of spot distance:} Energy resolved real space PL crosscuts for $y=0$ corresponding to the data points depicted in Fig.~2(g) of the main text. The dashed red lines marks the edges of the polariton branches (i.e., the bandgap around the BIC).}
    \label{supp_4}
\end{figure*}

\section{Experimental observation of the polarization vortex}
\begin{figure*}[hb]
    \centering
    \includegraphics[width=0.8\linewidth]{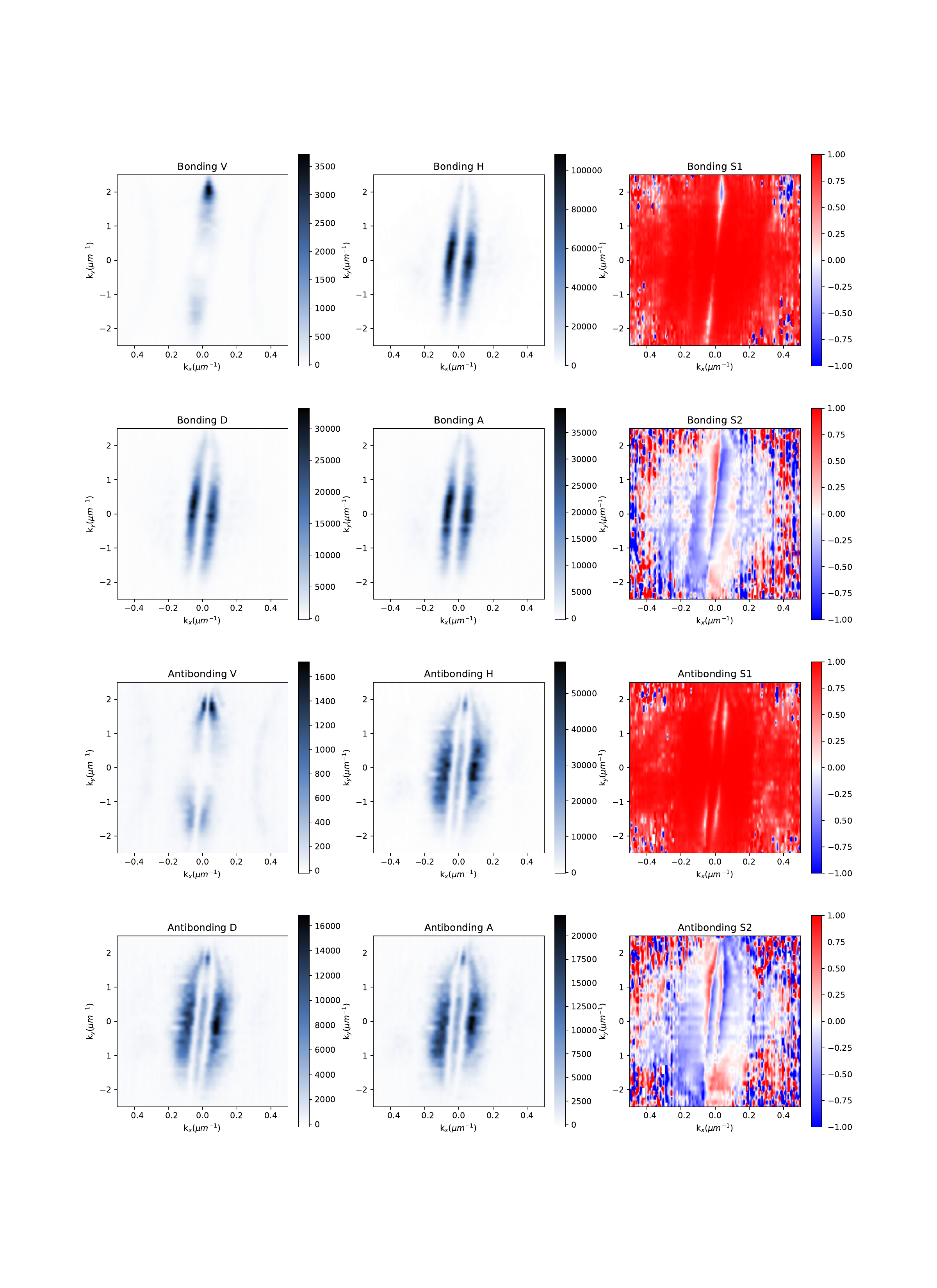}
    \caption{{\bf Experimental bonding and antibonding state polarization maps:} (a-f) H-V, D-A polarization resolved PL maps together with the resulting $S_1$ and $S_2$ Stokes parameters for the bonding state. (g-l) Identical characterisation for the antibonding state.}
    \label{supp_3}
\end{figure*}

Figure~\ref{supp_3} shows the raw data resulting in the polarization vortex reported in Fig.~3 of the main text. Four Fourier space PL maps corresponding to the four linear polarizations are shown in blue and yellow (V,H,D,A). We also plot the Stokes parameters $S_{1,2}$ which are defined
\begin{equation}
    S_{1,2} = \frac{I_{H,D} - I_{V,A}}{I_{H,D} + I_{V,A}}
\end{equation}
where $I_{H,V,A,D}$ are the intensities of emitted light of horizontal, vertical, diagonal, and antidiagonal linear polarization. The maps displayed here are slightly rotated with respect to the axis connecting the condensates. This is because of a tilting between the spectrometer slits and the corrugation direction of the grating. In the main text we represent the final maps rotated for the sake of clarity. The first 6 panels depict the bonding state on the main text showing the characteristic polarization rotation around the BIC state where the PL is mostly suppressed. The last 6 panels depict the antibonding state, where the polarization resolved PL maps reveal a clear double notch pattern. Again, in this instance, the two polarization vortices are located in the dark notches of the PL.

\section{Increasing the complexity of the potential landscape}
\begin{figure}[hb]
    \centering
    \includegraphics[width=0.4\linewidth]{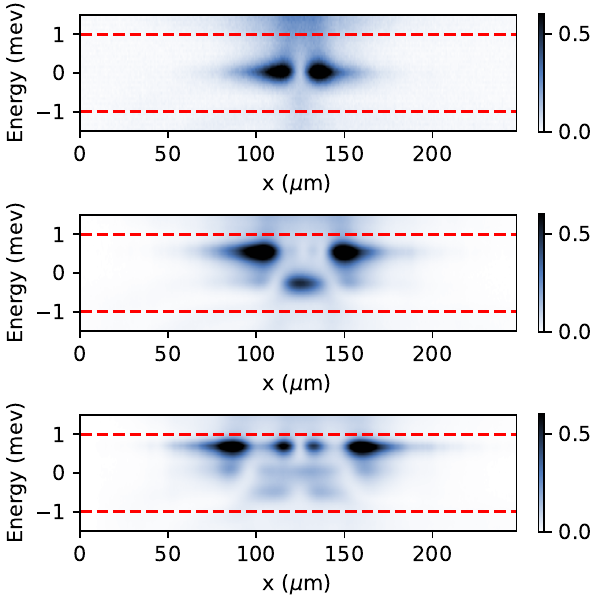}
    \caption{{\bf Multiple well potential:} Energy resolved real space PL crosscut for $y = 0$ direction showing single, double, and triple spot (trap) case for a inter-spot distance of 31 $\mu$m. The dashed red lines marks the edges of the polariton branches (i.e., the bandgap around the BIC). }
    \label{supp_2}
\end{figure}

Figure~\ref{supp_2} shows the energy resolved real-space photoluminescence (PL) cross sections for the 1,2 and 3 pump spots (i.e., single trap, double trap, and triple trap). Note that the observed confined states are contained within the bandgap (i.e., the states are not blueshifted into the upper dispersion branch) and exhibit alternating parity.

\section{Evanescent versus Ballistic coupling}

\begin{figure*}
    \centering
    \includegraphics[width=0.8\linewidth]{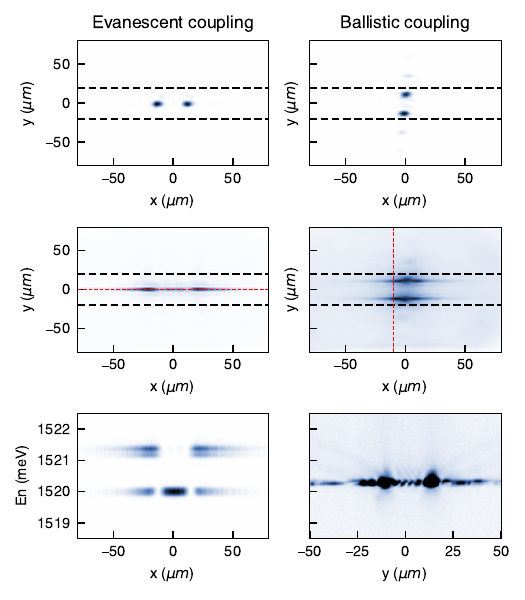}
    \caption{{\bf Coupling tuning by altering the relative angle between the condensates and the grating axis.:} Left column
denotes the studied evanescent coupling and the right column the case of ballistic coupling. Top row shows the real space non resonant laser excitation pattern, middle row the corresponding condensates emission. Bottom row shows the energy resolved crosscut along the red dashed line depicted in the middle row. The white lines mark approximately the grating edges. To note the different effect in case of evanescent coupling, where a splitting between bonding and antibonding states is clearly visible, while for ballistic coupling, on the same energy scale, no
splitting is present due to the weak interaction.}
    \label{supp_s10}
\end{figure*}


\neww{Since the dispersion in the direction perpendicular to the grating (i.e., parallel to the grooves) is still concave-up parabolic, the entire lower polariton branch has a saddle-like shape as reported in \cite{Ardizzone_Nature2022}. It is therefore possible, by simply adjusting the relative angle between the axis connecting the pump spots and the grating axis, to tune the BIC condensate coupling from negative-mass-evanescent to positive-mass-ballistic as we show in Figure \ref{supp_s10}. 

From this dataset, we can observe that our platform exhibits two distinct forms of coupling: evanescent coupling perpendicular to the grating grooves and ballistic coupling parallel to them. The nature of these two forms of coupling is drastically different, as depicted in Fig. \ref{supp_s10}(e) and \ref{supp_s10} (f), which show energy-resolved cross-sections in real space along the two coupling directions indicated by the red dashed lines in panels c and d.

In panel (e), the bonding and antibonding states can be discerned, with an energy splitting exceeding 1 meV, while in panel (f) the energy splitting results absent and of condensate signal match the energy of the uncoupled condensates, see Figure 2 (g) of the main text.

In absence of evanescent coupling, observing a luminescence crosscut orthogonal to the spot axis [see Figure \ref{supp_s10} (f)], the two condensates sitting on the top of a two blueshift hills can be spotted together with the high k polariton flow and the interference pattern demonstrating the phase locking of the two sites.

This set of measures allowed us to envisage a further path to extend the arrays beyond the 1D, maybe hybrid ballistic-evanescent coupled arrays of polariton condensates. In the manuscript, we focused on discussing the 1D physics along the grating direction for the sake of simplicity, where ballistic propagation is forbidden. The 2D picture is actually very rich and promising for future investigation.}

\section{Reciprocal space topological charge multiplicity in the multi-spot configuration}

\begin{figure*}
    \centering
    \includegraphics[width=0.8\linewidth]{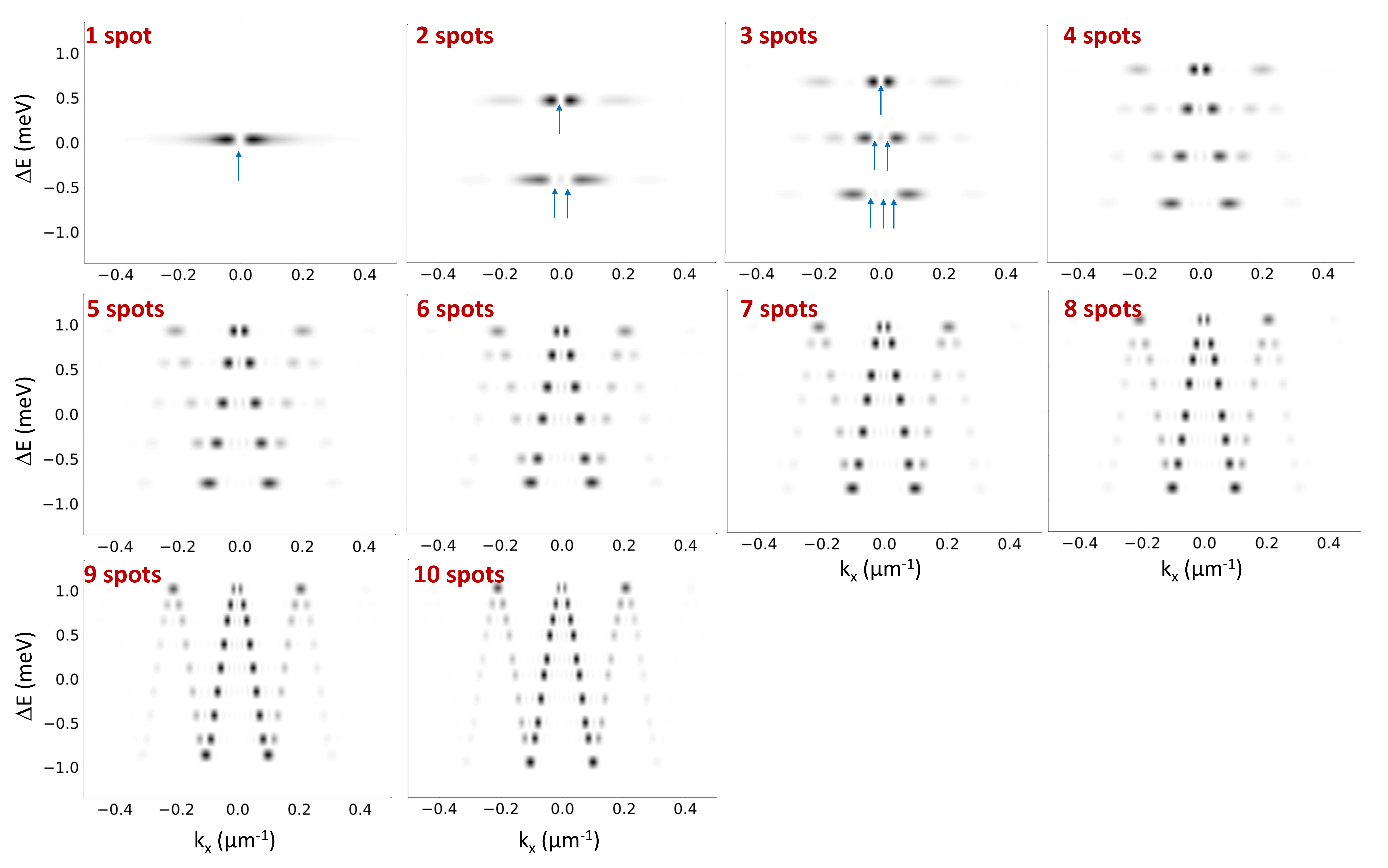}
    \caption{ \new{{\bf Theoretical calculation of the farfield intensity in k-space when the number of pumping spots is gradually increased from 1 to 10:} The pumping regime is the one so that the single spot pumping scheme will induce only a single confined state.  The notches in the vicinity of $k=0$ for the first three configurations are indicated by blue arrows.}}
    \label{multispot_vorticity}
\end{figure*}

\new{In the main text we discuss the topological charge multiplicity in the two spot configuration. Intuitively one would expect that introducing more and more sites in to the condensates chain, while keeping the same pumping regime, would lead to a linear increase of the charge multiplicity.
Interestingly this results to not be so straightforward.

In order to discuss the effect of the number of condensates on to the topological charge we have performed numerical simulations, of the far field photoluminescence, with the number of spots gradually increasing from 1 spot to 10 spots. The parameters are set in the tight binding limit: single bound state in a spot, which can weakly tunnel to the next. The results are shown in Figure \ref{multispot_vorticity}. These results indeed show that, in the case of 3 and 4 spots, the third and fourth-order states exhibit 3 and 4 notches in the vicinity of $k=0$, respectively. However, increasing the number of spots tends to clear away the non-zero signal between the notches in the excited states, and these high-order notches become barely visible for 5 and 6 spots. For 10 spots, we clearly observe the formation of the Bloch band that exhibits a single notch at $k=0$. }

\section{Heavy and light hole exciton transitions}

\new{Employing 20 nm QW results in a light hole (LH) transition relatively close to the heavy hole (HH). We have performed reflectivity measurement using a white light source to show the presence of these two excitons and their coupling with the optical modes of the grating. As can be seen in Fig. R2, the LH (approx 1.5355 eV) is 4.5 meV above the HH (at approx 1.531 eV) resonance and they couple differently with the TE and the TM modes.

In our case, it is reasonable to assert that the presence of the LH does not affect the TE polarized branch. It is well known in the literature that the oscillator strength associated with the LH transition is at least three times lower compared to the HH transition [M. N. Bataev et. al, \cite{bataev_heavy-hole--light-hole_2022}]
}

\begin{figure}
    \centering
    \includegraphics[width=0.6\linewidth]{images/final images/figS6.pdf}
    \caption{\new{\textbf{Heavy and light hole exciton transition: }Polarized reciprocal space white light reflectivity spectra for 2 differently detuned structures reported in section S2 corresponding to a pitch of 240 (a,b) and 241 (c,d) nm respectively.}}
    \label{heavy-light}
\end{figure}

\section{Exciton-BIC detuning}

\begin{figure*}[t]
    \centering
    \includegraphics[width=0.6\linewidth]{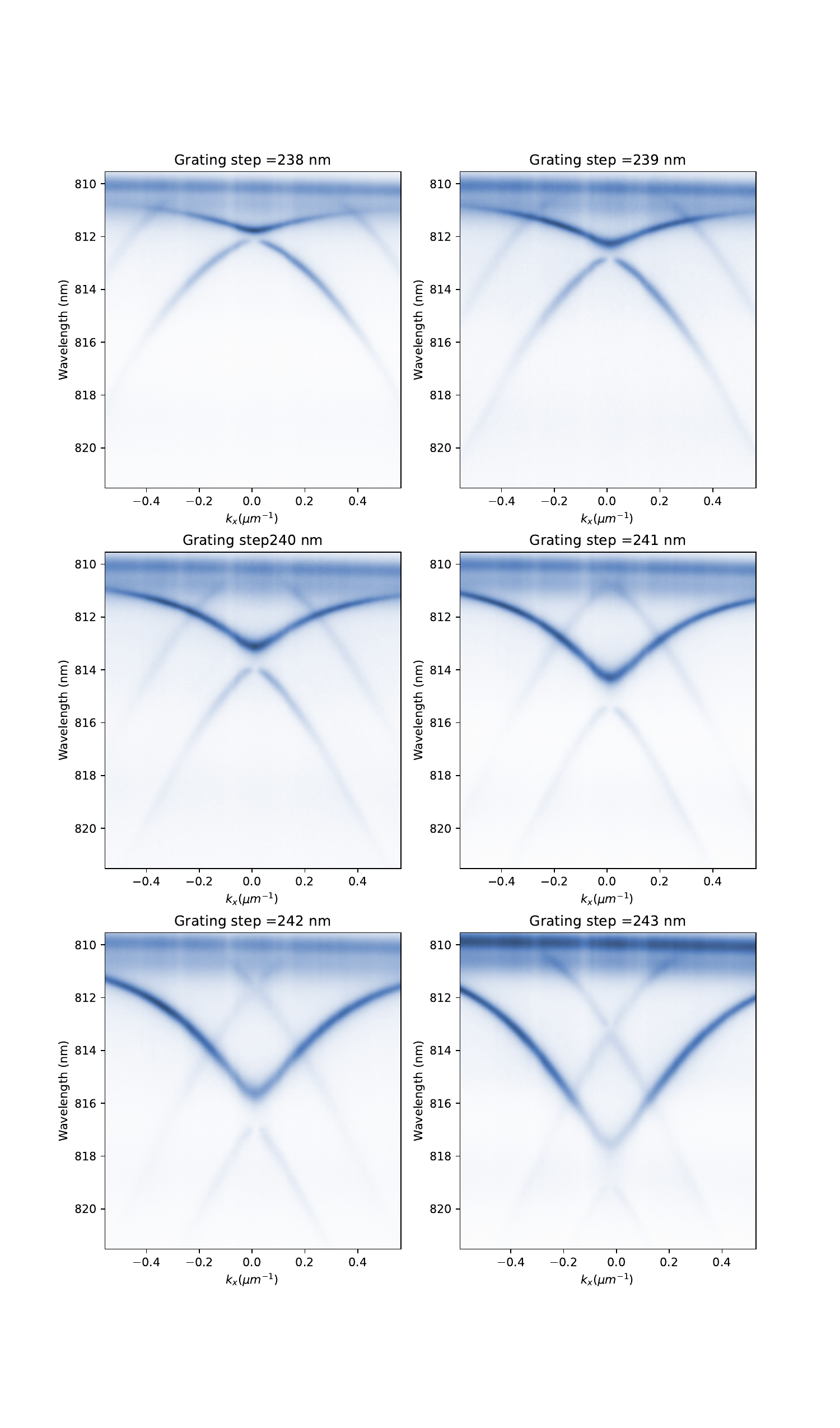}
    \caption{{\bf Exciton-BIC detunings:} Experimental energy momentum dispersion crosscut along $k_{y} =0$ for all the six detunings available on to the sample.}
    \label{supp_1}
\end{figure*}

As already mentioned in the Methods section of the main text the sample is composed of an active GaAs/AlGaAs waveguide in strong coupling regime. Figure~\ref{supp_1} shows the anti-crossing behaviour between the photon and exciton dispersion with a clear Rabi splitting. 
In polaritonic systems a second crucial parameter in determining the polariton properties is the exciton-photon detuning which, together with the Rabi energy, determines the Hopfield coefficients of the polariton state~\cite{Deng_RMP2010}.

Differently from a planar microcavity where the optical mode is a Fabry-P\'erot mode and the energy can be tuned by adjusting the cavity thickness, in a grated waveguide the photon group velocity is fixed by the refractive index of the material. In a polariton BIC system the energy separation between the photonic BIC and the exciton is determined by the grating step. In Fig.~\ref{supp_1} we report 6 different exciton-BIC detunings for grating steps ranging from 238 to 243 nm.

The experimental results reported in the main text have been collected on the second the third and the forth structure depending on the need of the specific experiment. We used the third structure for the results in Fig.~1 in the main text; structure 4 for measuring the molecule energy splitting and the polarization vortex (Fig.~2 and~3 in the main text); and finally the second structure for observing the Bloch band formation reported in Fig.~4 in the main text.

\section{BIC quality factor estimation in a finite active structure}

\new{In this section we address the effect of the finite size and the exciton coupling onto the quality factor of the polariton BIC, Figure \ref{quality_factor}. 
We first calculate the quality factor of the BIC branch for the infinite structure for the purely photonic modes (blue circles) 
The simulation has been performed via RCWA (Rigourous Coupled Wave Analysis) method. It shows that the quality factor of the passive (i.e., without exciton) structure diverges approaching $k=0$, as expected for the symmetry protected mode. In presence of an oblique wavevector ($k\neq0$), the mirror symmetry is broken and the photonic quality factor becomes finite. As shown in the figure \ref{quality_factor}, the dependance on $k$ of the passive quality factor $Q_{passive}(k)$ can be nicely fitted by a simple quadratic model\cite{dang_realization_2022}:  
\begin{equation}
Q_{passive}(k)= Q_{\infty} +\frac{\alpha}{k^2}. \label{eq:Q_factor_passive}
\end{equation}
Where $\alpha/k^2$ represents the decrease of radiative quality factor of the symmetry protected BIC band when oblique wave-vector is introduced, and the quality factor at high wavevector is given by $Q_{\infty} = 5000$.

For a finite size grating, the momentum space is now quantified by $ \delta_k=\pi/L=0.01\, \mu m^{-1}$ where $L=300\,\mu m$ is the lateral size of the grating. As a consequence, the photonic BIC becomes a quasi-BIC, with quality factor given by $Q_{passive}^{BIC}=Q_{passive}(k=\delta_k)$ [see Zhou et al,  \cite{zhou_increasing_2023}]. Using this simple rule, we extract from the previous result that for a finite structure, we expect a quality factor of $Q_{passive}^{BIC}\approx 2\cdot10^5$ (vertical dashed line in Fig. \ref{quality_factor}).

When photonic BIC is strongly coupled to excitonic resonance to form polariton BIC, the active quality factor can be modeled by (see Dang et al., \cite{dang_realization_2022}):
\begin{equation}
    \frac{1}{Q_{active}^{BIC}}=\frac{|X|^2}{Q_X}+\frac{|C|^2}{Q_{passive}^{BIC}}. \label{eq:Q_factor}
\end{equation}
where $Q_X$ is the non-radiative quality factor of the excitonic resonance, while $|X|^2=0.35$ and $|C|^2=0.65$ are the excitonic and photonic fraction of the polaritonc state respectively.  Since the lifetime of BIC polaritons exceeds 300 ps\,\cite{Ardizzone_Nature2022}, we can assume a lifetime 1 ns for heavy hole excitons. This corresponds to $\gamma_X\approx0.6\cdot10^{-3}$ meV, thus $Q_X\approx 2.3 \cdot10^{7}$. Therefore, using the expression \eqref{eq:Q_factor}, we estimate $Q_{active}^{BIC}\approx 3\cdot10^5$. We note that for the active and finite structure, the quality factor is limited by the finite size of the photonic grating and not by the non-radiative lifetime of exciton.

\begin{figure*}
    \centering
    \includegraphics[width=0.6\linewidth]{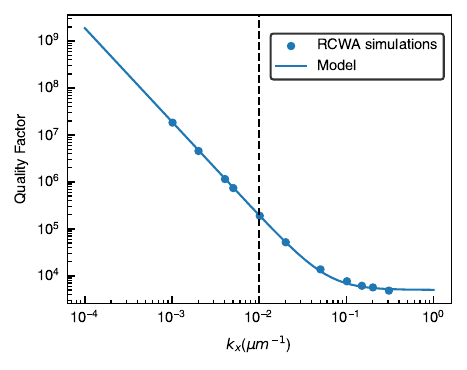}
    \caption{ \new{{\bf The quality factor as the function of the wavevector:} Blue circles are results obtained from numerical simulations of infinite photonic grating. Solid line is the fitting using the theoretical model \eqref{eq:Q_factor_passive}. Vertical black dashed line corresponds to the $\delta$k of the quantification in momentum space due to the finite size of the grating}}
    \label{quality_factor}
\end{figure*}
}

\section{Experimental linewidth of the polaritonic BIC}

\begin{figure*}
    \centering
    \includegraphics[width=0.55\linewidth]{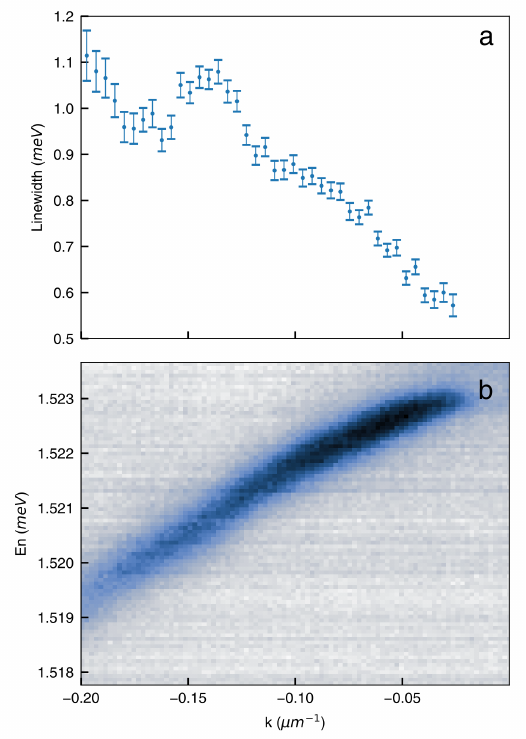}
    \caption{ \new{{\bf Linewidth of the polaritonic BIC branch approaching $k=0$ for the 240 nm pitch grating:}  a) Full width at half maximum (FWHM) of the polaritonic mode measured for different in-plane momenta, the error bars are \neww{the confidence interval of the FWHM parameter obtained fitting the luminescence profiles with a Gaussian function} b) corresponding photoluminescence.
}}
    \label{linewidth}
\end{figure*}

\new{In Figure \ref{linewidth}\,a we present the linewidth extracted from the BIC branch as a function of the in-plane momenta, and in Figure \ref{linewidth}\,b the corresponding PL. This figure clearly demonstrates that, unlike the classical microcavity platform, the linewidth of the BIC state cannot be straightforwardly determined from classical energy-resolved photoluminescence due to the lack of light around $k=0$.

Furthermore, the dark nature of the BIC state also prevents resonant probing of its lifetime, because a resonant beam would not be injected into the structure. These reasons explain why, in our previous work [Ardizzone et al. \cite{Ardizzone_Nature2022}], we employed a propagation method, extensively described in the referenced paper above, to estimate the polariton lifetime close to the BIC state which gave 300 ps (a huge improvement from conventional polariton cavities).
}

\section{Exciton reservoir extension}
\begin{figure*}[t]
    \centering
    \includegraphics[width=0.5\linewidth]{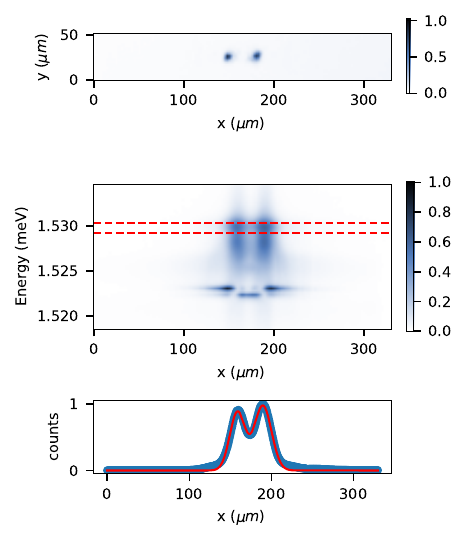}
    \caption{{\bf Exciton reservoir extension:} (a) Real space distribution for the excitation laser focused on the waveguide consisting in two 6.5 $\mu$m FWHM Gaussian spots separated by 31 $\mu$m distance. (b) Energy resolved PL crosscut corresponding to the excitation scheme in panel (a). (b) Resulting PL profile, the blue points represent the experimental points corresponding to the region between the red dashed line in panel (b). The red continuous line is a double Gaussian fitting with separation distance $31~\mu$m and a FWHM of $22~\mu$m.}
    \label{supp_5}
\end{figure*}

As discussed in the main text, a FWHM $6.5~\mu$m laser spot results in a wider effective potential well. To characterize the increased width of the effective optical traps with respect to the laser beam width in Fig.~\ref{supp_5}(a) we show our laser excitation scheme consisting of two $6.5~\mu$m wide laser spots separated by $31~\mu$m. In panel (b) the corresponding energy resolved profile is shown. Here the contribution from the exciton can be distinguished at 1530 meV together with the PL contribution from the upper BIC branch and the bonding and antibonding states. In Fig.~\ref{supp_5}(c) we obtain a $22~\mu$m FWHM of fitting the resulting PL with two Gaussians. This demonstrates the extension of the exciton reservoir, due to diffusion, way beyond the laser excited region.

\bibliography{biblio}